# The Heavy Quark Masses from Quarkonia[*][†]

Aida X. El-Khadra[a] and B. P. Mertens[b]

[a]Physics Department, Ohio State University, 174 W 18th Ave, Columbus, Ohio, USA

[b]Enrico Fermi Institute and Department of Physics, University of Chicago, Chicago, Illinois, USA

We report on the status of the determination of the heavy quark masses from our calculation of the quarkonia spectra. All sources of systematic errors that enter the quark mass determination are accounted for. We explicitly keep $ma \neq 0$ in the perturbative calculation relating the bare lattice mass to a renormalized mass. Our results are still preliminary.

## 1. INTRODUCTION

Quarkonia, mesons containing a heavy quark and anti-quark, are at present the best understood hadronic systems. As has been argued by Lepage [1], quarkonia are also the easiest systems to study with lattice QCD, and systematic errors can be analyzed using potential models. Control over systematic errors in turn allows the extraction of Standard Model parameters from the quarkonia spectra.

We report on the status of the charm-quark mass determination from the $c\bar{c}$ spectrum using the Fermilab action [2]. This action is a generalization of previous approaches, which encompasses the non-relativistic limit for heavy quarks as well as Wilson's relativistic action for light quarks. Lattice-spacing artifacts are analyzed for quarks with arbitrary mass. The determination of the strong coupling from the $c\bar{c}$ and $b\bar{b}$ spectra has already been presented elsewhere [3].

## 2. THE SPECTRUM

Figure 1 shows the $c\bar{c}$ spectrum calculated with the Fermilab action in the quenched approximation [4]. The lattice spacing is set with the $h_c$−1S splitting (see Refs. [3,4] for details on the numerical calculations).

Let $E$ denote the energy of the quarkonium

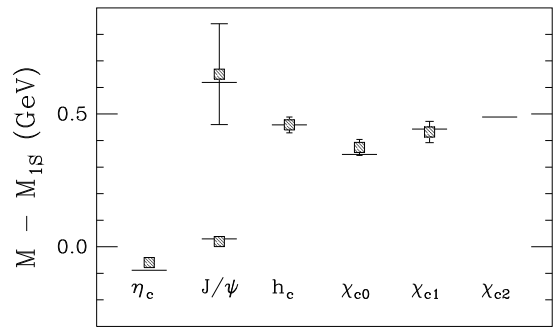

Figure 1. The charmonium spectrum in comparison with experiment.

state, extracted from the exponential fall-off. The rest mass, $M_1 = E(\mathbf{p}=0)$, and the kinetic mass, $M_2 = [d^2 E(\mathbf{p})/d\mathbf{p}^2]^{-1}_{\mathbf{p}=0}$, are unequal [2] for the actions used in our numerical calculations. This is acceptable in non-relativistic systems if the bare lattice mass is tuned such that the kinetic mass, $M_2$, equals the desired ground-state mass. Fig-

Table 1
Comparison of higher order lattice spacing errors for the 1S state and the 1P−1S splitting.

| $a^{-1}$ (GeV) | $c$ | 1S (MeV) | 1P−1S (MeV) |
|---|---|---|---|
| 2.4 | 1.4 | +30 | − 4 |
|     | 0.0 | +50 | −30 |
| 1.8 | 1.4 | +40 | − 7 |
|     | 0.0 | +70 | −40 |

---

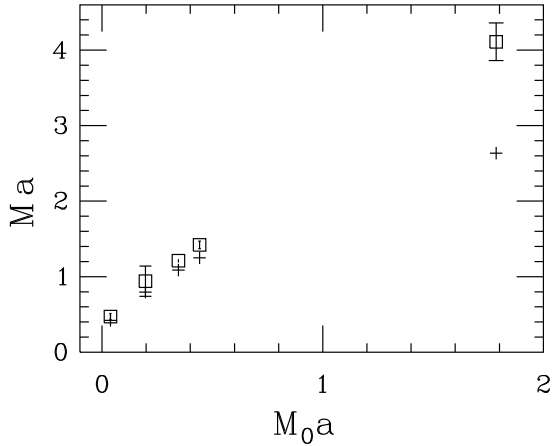

Figure 2. Numerical results for $M_1 a$ (+) and $M_2 a$ (□) vs. $M_0 a$.

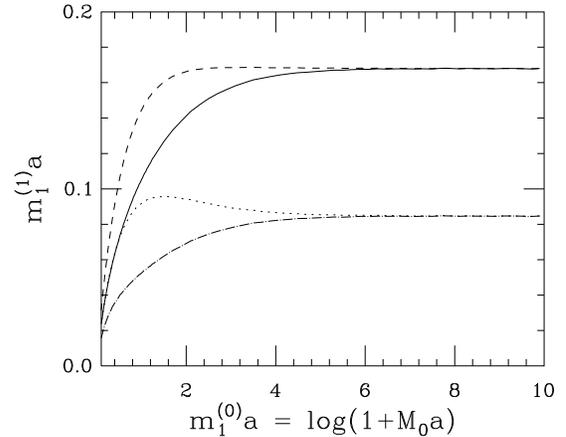

Figure 3. $m_1^{(1)} a$ as function of $m_1^{(0)} a$ for the Wilson action (solid), improved action with $c = 1.4$ (long dashed) and with tadpole improvement for both actions (dash-dotted, short dashed).

ure 2 shows the masses $M_1$ and $M_2$ as functions of the bare lattice mass $M_0 = \frac{1}{2\kappa} - \frac{1}{2\kappa_c}$ in lattice spacing units for the $24^3 \times 48$ lattice at $\beta = 6.1$ [4]. The effect of higher order lattice spacing errors on the spectrum can be estimated using potential models. Table 1 compares the effect for the spin-averaged 1S state and the 1P–1S splitting in the $c\bar{c}$ spectrum at several lattice spacings and for the Wilson and the improved ($c = 1.4$) action [4]. Richardson potential model wave functions were used for the estimate.

## 3. THE SELF ENERGY

The calculation of the self energy with $m_0 a \neq 0$ has been outlined in [5], where preliminay results at one-loop for the Wilson action were shown. In order to distinguish quark and meson masses, we shall denote meson masses by $M_{1,2}$ and quark masses by $m_{1,2}$. The perturbative expansion for $m_1$ is written as

$$m_1 a = \log(1 + M_0 a) + g^2 m_1^{(1)} a + \mathcal{O}(g^4) \ . \quad (1)$$

Figure 3 shows the one-loop term, $m_1^{(1)}$, for the Wilson and improved ($c = 1.4$) actions as a function of $m_1^{(0)} a = \log(1 + M_0 a)$ with and without tadpole improvement (using the plaquette) [8]. The corresponding effective scales for the coupling, $q^*$, as defined in Ref. [6] (eq. (19)) are shown in Figure 4.

## 4. RESULTS

The determination of the charm quark mass proceeds analogously to the $b$ quark mass determination of Ref. [7]. The binding energy is obtained from $E_{\text{bind}} = M_1 - 2m_1$, which together with the experimentally observed spin-averaged mass of the 1S state, $M_{1S}^{\text{exp}} = 1/4(M_{\eta_c} + 3M_{J/\psi})$, determines the charm quark pole mass as

$$m_{\text{ch}}^{\text{pole}} = \frac{1}{2}(M_{1S}^{\text{exp}} - E_{\text{bind}}) \ . \quad (2)$$

Alternatively, after tuning the bare quark mass, such that $M_2 = M_{1S}^{\text{exp}}$, the pole mass can be obtained from:

$$m_{\text{ch}}^{\text{pole}} = m_2 = Z_m m_2^{(0)} \ , \quad (3)$$

with $Z_m = 1 + g^2 Z_m^{(1)}$ at one-loop. Preliminary results for $Z_m^{(1)}$ exist. Once the calculation is completed [8], the comparison of eqs. (2) and (3) will give an important consistency check.

Of course, as always, all systematic errors arising from the lattice QCD calculation need to be under control for a phenomenologically interesting result. In particular, the systematic error introduced by the omission of sea quarks has to be removed. The short-distance corrections that introduced the dominant uncertainty to the $\alpha_s$ determination from quarkonia [3] are absent for the pole mass determination, because this effective mass does not run for momenta below it's mass. This argument has been tested from first principles in Ref. [7] for the $b$ quark pole mass with consistent results.

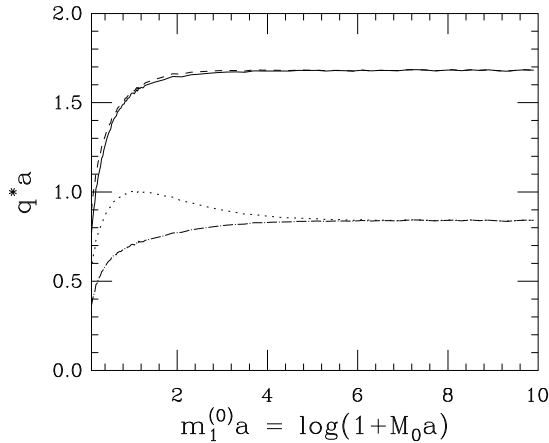

Figure 4. $q^*$ for $m_1^{(1)}$ as function of $m_1^{(0)}a$ for the Wilson action (solid), improved action with $c = 1.4$ (long dashed) and with tadpole improvement for both actions (dash-dotted, short dashed).

The statistical errors are dominated by the 1P–1S splitting which determines $a^{-1}$, and are $< 10\%$ of the binding energy. An estimate of the higher order lattice spacing errors to the spin-averaged spectrum is listed in table 1. Finite volume errors are negligible.

The dominant systematic error comes from the unknown higher order corrections to eq. (1) (and eq. (3)); a preliminary estimate is $100-200$ MeV. Our preliminary result is $m_{\text{ch}}^{\text{pole}} = 1.5(2)$ GeV.

The $\overline{\text{MS}}$ mass for the charm quark has been determined from a compilation of $D$ meson calculations in the quenched approximation [9]. However not all systematic errors were considered. In particular, sea-quark effects cannot, in this case, be estimated phenomenologically, leaving this systematic error uncontrolled. Once our analysis for the $\overline{\text{MS}}$ mass is completed, we will be able to compare our result with that of Ref. [9].

The analysis of the $b\bar{b}$ spectrum for a determination of the $b$ quark pole (and $\overline{\text{MS}}$) mass is also in progress.

As already pointed out in Ref. [7], the heavy quark mass determination is limited by our lack of knowledge of the higher order perturbative terms. The extension of the self energy calculation to two-loops will require tools different from those used in the one-loop calculation [10].

## ACKNOWLEDGEMENTS

The perturbation theory is being done in collaboration with A. Kronfeld and P. Mackenzie [8]. The quarkonium spectrum is calculated in collaboration with G. Hockney, A. Kronfeld, P. Mackenzie, T. Onogi, and J. Simone [4]. B.P.M. is supported by the U. S. Department of Energy under Grant No. DE-FG02-90ER40560.

## REFERENCES

1. P. Lepage, Nucl. Phys. **B** (Proc. Suppl.) **26** (1992) 45; B. Thacker and P. Lepage, Phys. Rev. **D43** (1991) 196; P. Lepage and B. Thacker, Nucl. Phys. **B** (Proc. Suppl.) **4** (1988) 199.
2. A. El-Khadra, A. Kronfeld and P. Mackenzie, Fermilab PUB-93/195-T.
3. A. El-Khadra, G. Hockney, A. Kronfeld and P. Mackenzie, Phys. Rev. Lett. **69** (1992) 729; A. El-Khadra, Nucl. Phys. **B** (Proc. Suppl.) **34** (1994) 141.
4. A. El-Khadra, G. Hockney, A. Kronfeld, P. Mackenzie, T. Onogi and J. Simone, Fermilab PUB-94/091-T, in preparation.
5. A. Kronfeld and B. Mertens, Nucl. Phys. **B** (Proc. Suppl.) **34** (1994) 495;
6. P. Lepage and P. Mackenzie, Phys. Rev. **D48** (1992) 2250.
7. C. Davies, *et al.*, Phys. Rev. Lett. **73** (1994) 2654.
8. A. El-Khadra, A. Kronfeld, P. Mackenzie and B. Mertens, in preparation.

9. C. Allton, *et al.*, CERN-TH.7256/94, hep-ph/9406263.
10. P. Mackenzie, this volume.